\documentclass[twocolumn,showpacs,preprintnumbers]{revtex4}
\usepackage{graphicx}
\usepackage{dcolumn}
\usepackage{bm}
\usepackage{subfigure}
\begin{document}
\newcommand{\ds}{\displaystyle}
\newcommand{\be}{\begin{equation}}
\newcommand{\en}{\end{equation}}
\newcommand{\bea}{\begin{eqnarray}}
\newcommand{\ena}{\end{eqnarray}}
\title{Phantom evolving wormholes with big rip singularities}
\author{Mauricio Cataldo}
\altaffiliation{mcataldo@ubiobio.cl} \affiliation{Departamento de
F\'\i sica, Facultad de Ciencias, Universidad del B\'\i o-B\'\i o,
Avenida Collao 1202, Casilla 5-C, Concepci\'on, Chile.}
\author{Paola Meza}
\altaffiliation{paolameza@udec.cl} \affiliation{Departamento de F\'{\i}sica, Universidad de Concepci\'{o}n,\\
Casilla 160-C, Concepci\'{o}n, Chile.}
\date{\today}
\begin{abstract}
We investigate a family of inhomogeneous and anisotropic
gravitational fields exhibiting a future singularity at a finite
value of the proper time. The studied spherically symmetric
spacetimes are asymptotically Friedmann-Robertson-Walker at spatial
infinity and describe wormhole configurations filled with two matter
components: one inhomogeneous and anisotropic fluid and another
isotropic and homogeneously distributed fluid, characterized by the
supernegative equation of state $\omega=p/\rho < -1$. In previously
constructed wormholes, the notion of the phantom energy was used in
a more extended sense than in cosmology, where the phantom energy is
considered a homogeneously distributed fluid. Specifically, for some
static wormhole geometries the phantom matter was considered as an
inhomogeneous and anisotropic fluid, with radial and lateral
pressures satisfying the relations $p_{r}/\rho<-1$ and $p_{_l} \neq
p_r$, respectively. In this paper we construct phantom evolving
wormhole models filled with an isotropic and homogeneous component,
described by a barotropic or viscous phantom energy, and ending in a
big rip singularity. In two of considered cases the equation of
state parameter is constrained to be less than $-1$, while in the
third model the finite-time future singularity may occur for
$\omega<-1$, as well as for $-1 < \omega \leq 1$.

\vspace{0.5cm} 
\end{abstract}
\smallskip
\maketitle 
\section{Introduction}
Recent astrophysical observations indicate that our Universe is
currently in accelerating expansion~\cite{Acceleration,Copeland}.
This discovery has stimulated an intensive study of models where a
large number of possible cosmological mechanisms have been proposed
to explain the origin of the current acceleration. In the framework
of general relativity, a quantitative analysis shows that a
mysterious component of energy, dubbed dark energy, is responsible
for the origin of this cosmic phenomenon and it starts dominating
the matter content dynamics only at recent times, being irrelevant
at earlier stages of the evolution. Many dark energy models have
been proposed to solve this fundamental problem of cosmological
physics, such as $\Lambda$-CDM model~\cite{LambdaCDM}, quintessence
models~\cite{Quintessence,Copeland}, k-essence models~\cite{k
essence}, phantom models~\cite{phantom models}, etc.

These new advances in cosmology allow us to consider new type of
singularities, besides already considered standard singularities
such as Big-Bang and Big-Crunch~\cite{Big Bang}. This non-standard
type of singularities occurs at a finite value of cosmological time
and are included in the following descriptive classification: the
singularity is a Big Rip when the scale factor, energy density and
pressure go to infinity in a finite proper time, and is a sudden
singularity when at a finite value of time and scale factor,
curvature or one of its higher derivatives blow up~\cite{Future
Singularities}.

The main motivation of this type of singularities comes from phantom
cosmological models. In the framework of Friedmann-Robertson–Walker
(FRW) cosmologies filled with two matter contents and dominated by a
phantom type fluid, satisfying a non-dissipative barotropic equation
of state, this new type of singularities can be completely
classified in the following four types: type I for a Big Rip
singularity; type II for a sudden singularity defined by a finite
energy density and diverging pressure; type III for diverging energy
density and pressure at a finite value of the scale factor, and the
type IV for finite curvature components and diverging higher
derivatives of $H$~\cite{Nojiri}.

On the other hand, since the pioneering works by Morris and
Thorne~\cite{Morris}, the study of wormholes has become one of the
most popular and intensively studied topics in relativistic physics,
where most of the efforts are directed to study Lorentzian
wormholes, in the framework of classical general relativity,
sustained by an exotic matter with negative energy density. These
models include both static~\cite{StaticWH} and evolving relativistic
versions~\cite{EvolvingWH}, sustained by a single fluid component.
The interest has been mainly devoted to traversable wormholes, which
have no horizons, allowing two-way passage through
them~\cite{LoboVisser}.

For static wormholes the fluid requires the violation of the null
energy condition (NEC), while in Einstein gravity there are
nonstatic Lorentzian wormholes which do not require WEC violating
matter to sustain them. Such wormholes may exist for arbitrarily
small or large intervals of time~\cite{KarSahdev} or even satisfy
the dominant energy condition (DEC) in the whole
spacetime~\cite{Maeda,Cataldo15}. One can consider also dynamic
wormhole spacetimes filled with two fluids, just like it is required
in cosmology where such two-fluid models are widely considered today
in order to explain the observed accelerated expansion of the
Universe~\cite{Pavon2}.

Wormhole spacetimes filled with phantom type matter were considered
before~\cite{Phantom WH}. Specifically, spherically symmetric static
wormholes were studied, sustained by a phantom type matter (or
super-quintessence) with anisotropic pressure. In these static
models the notion of the phantom energy is used in a more extended
sense than in cosmology since, strictly speaking, the phantom matter
is a homogeneously distributed fluid, and for these non-dynamic
wormhole models an inhomogeneous and anisotropic matter component is
used, having for the radial and lateral pressures $p_{r}<-1$ and
$p_{_l} \neq p_r$, respectively~\cite{Lobo1}. This type of extended
phantom-like matter was also used in dynamical wormhole
models~\cite{Cataldo1}.

The time evolution of wormhole geometries in a Friedmann universe
exhibiting a Big Rip singularity were previously studied. However
the literature on this topic is not extensive. For example, in
Ref.~\cite{Nabulsi} the author considers some accelerated
higher-dimensional cosmologies with a traversable static wormhole,
dominated by a time-dependent cosmological constant, and ending at a
Big Rip. The studied Big Rip solutions have an exponential scale
factor. In addition, the authors of Ref.~\cite{Israel} consider two
different wormhole models, modeled by a thin spherical shell
accreting the phantom fluid.

It is interesting to note that in this context, it has recently
proposed that it is possible that the universe could avoid the big
rip singularity with the occurrence of a big
trip~\cite{P.F.Gonzalez-Diaz,Faraoni}, which is a cosmological event
that may appear during the evolution of a wormhole embedded in a FRW
universe approaching the Big Rip singularity. In this case, the
wormhole accreting phantom matter expands faster than the background
FRW universe, and the radius of the wormhole throat diverges before
the Big Rip is reached. In this scenario, the wormhole engulfs the
entire universe, which will re–appear from the other wormhole
throat~\cite{Israel}.

In the present paper we intend to study evolving wormholes filled
with two matter components, where one of them is an isotropic
homogeneously distributed phantom fluid characterized the
supernegative equation of state $\omega=p/\rho < -1$, and presenting
a future singularity at a finite value of the proper time.

The organization of the paper is as follows: In Sec. II we present
the dynamical field equations for wormhole models with a matter
source composed of an ideal isotropic cosmic fluid and an
anisotropic and inhomogeneous one. In Sec. III expanding wormholes
filled with a barotropic dark energy and phantom fluids are studied.
We discuss explicit models ending at a finite-time future
singularity. In Sec. IV viscous expanding wormholes are discussed.
Models that may evolve to a finite-time future singularity are
considered, and in Sec. V we conclude with some remarks.

\section{Field equations}
In this paper we shall make use of some previously obtained results
by one of the authors in Ref.~\cite{Cataldo15}. Let us state the
main result obtained in that paper, concerning with solutions
containing two fluids and admitting spherical symmetry in the
framework of the Einstein gravity theory.

Taking the metric
\begin{eqnarray}\label{evolving wormhole15}
ds^2=-dt^2+  a(t)^2 \left( \frac{dr^2}{1-kr^2-\frac{b(r)}{r}}+r^2 d
\Omega^2 \right),
\end{eqnarray}
in comoving coordinates, filled with the anisotropic and
inhomogeneous fluid $\rho_{_{in}}(t,r)$, and the isotropic and
homogeneous fluid $\rho(t)$, the Einstein equations are given by
\begin{eqnarray}\label{00}
3 H^2+\frac{3k}{a^2}+\frac{b^{\prime}}{a^2 r^2}=\kappa \rho_{_{in}}(t,r)+\kappa \rho(t)+\Lambda,  \\
\label{rr} -  \left( 2\frac{\ddot{a}}{a}+ H^2 +\frac{k}{a^2}
\right)- \frac{b}{a^2
r^3} =\kappa p_r(t,r)+\kappa p(t)-\Lambda, \\
\label{thetatheta} -  \left(2 \frac{\ddot{a}}{a}+ H^2 +\frac{k}{a^2}
\right) + \frac{b-r b^{\prime}} {2 a^2 r^3}=\kappa
p_{_l}(t,r)+\kappa p(t) -\Lambda, \label{22}
\end{eqnarray}
where $d \Omega^2=d\theta^2+sin^2 \theta d \varphi^2$, $\kappa=8 \pi
G$; $\Lambda$ is the cosmological constant, $a(t)$ is the scale
factor, $k=-1,0,1$; $H=\dot{a}/a$; and an overdot and a prime denote
differentiation $d/dt$ and $d/dr$ respectively.

In this case the $4$-velocity of the fluids is given by the timelike
vector $u^{\alpha}=(1,0,0,0)$, and the radial and tangential
pressures obey the barotropic state equations
\begin{eqnarray} \label{FRWC00}
p_r(t,r)=\omega_r \, \rho_{_{in}}(t,r), \nonumber \\
p_{_l}(t,r)=\omega_{_l} \, \rho_{_{in}}(t,r),
\end{eqnarray}
with constant state parameters $\omega_r$ and $\omega_{_l}$.

Note that the essential characteristics of a wormhole geometry are
encoded in the spacelike section of the metric~(\ref{evolving
wormhole15}). It is clear that this metric becomes a zero-tidal
force static wormhole if $a(t) \rightarrow const$, and as $b(r)
\rightarrow k r^3$ it becomes a flat FRW metric for $k=0$, a closed
FRW metric for $k=1$, and an open FRW metric for $k=-1$.

It can be shown that Eqs.~(\ref{evolving wormhole15})-(\ref{22}) may
be rewritten in the form~\cite{Cataldo15}
\begin{eqnarray}\label{INHOMOGENEOUS FRW}
ds^2=dt^2  \nonumber \\ -a(t)^2 \left( \frac{dr^2}{1-k r^2+\kappa \,
C \, \omega_r \, r^{-1-1/\omega_r}}+ r^2 d\Omega^2\right),
\end{eqnarray}
\begin{eqnarray}\label{friedmann equation}
3 H^2+\frac{3k}{a^2} &=& \kappa \rho(t)+\Lambda, \\
\dot{\rho}+3H (\rho+p) &=& 0, \label{FRWCEq}
\end{eqnarray}
where $C$ is an integration constant, and the inhomogeneous and
anisotropic cosmic fluid is given by
\begin{eqnarray}
p_{_r}&=&\omega_r \, \rho_{_{in}}, \\
p_{_l}&=&-\frac{1}{2}(1+\omega_r) \, \rho_{_{in}}, \\
\rho_{_{in}}(t,r)&=&\frac{C \, r^{-3-1/\omega_r}}{ a^{2}(t)}.
\label{rhoC Final}
\end{eqnarray}
Here, the constraint $\omega_r+2\omega_{_l}+1=0$ for
Eqs.~(\ref{FRWC00}) was used. From these expressions we conclude
that if $\omega_{_r}>0$ or $\omega_{_r}<-1$ the obtained
gravitational configurations are asymptotically FRW solutions at
spatial infinity.

In conclusion, the main result of the Ref.~\cite{Cataldo15} is that
the evolution of the scale factor $a(t)$ in the
metric~(\ref{INHOMOGENEOUS FRW}) is governed by the standard
Friedmann equations~(\ref{friedmann equation}) and~(\ref{FRWCEq}),
and it is determined by the fluid $\rho(t)$. This matter component
may be in principle an ideal barotropic fluid or any other cosmic
fluid satisfying the requirements of isotropy and homogeneity.

\section{Expanding Wormhole Universes filled with a barotropic dark energy and phantom fluids}
In this section we shall consider that the isotropic and homogeneous
matter component is described by a barotropic phantom energy
$\rho(t)$ with an equation of state of the form
\begin{eqnarray}\label{SEPF}
p(t)=\omega \rho(t),
\end{eqnarray}
where the constant state parameter $\omega$ satisfies the constraint
$\omega<-1$. In general Eq.~(\ref{SEPF}) allows us to consider a
barotropic matter component describing standard matter for $\omega
\geq 0$, a dark energy fluid for $-1<\omega<-1/3$, and
super-quintessence for $\omega< -1$, just like it is defined in
cosmology.

From now on in this section, we shall consider solutions with
$k=\Lambda=0$ and $\kappa=8 \pi G=1$. Thus in this case, from
Eqs.~(\ref{friedmann equation}) and~(\ref{FRWCEq}), we obtain that
the scale factor is given by $a(t)=D(F+(3/2) (\omega+1) t)^{2/(3
(\omega+1))}$, where $D$ and $F$ are constants of integration. This
scale factor we shall rewrite as
\begin{eqnarray}\label{SFsinxi}
a(t)=a_{_{0}} \left(1+\frac{3}{2} H_{_0}(\omega+1) t
\right)^{2/(3(\omega+1))},
\end{eqnarray}
and the energy density in the form
\begin{eqnarray}\label{EDh}
\rho(t)=\frac {\rho_{_{0}}}{ \left( 1+\frac{3}{2} H_{_0} (\omega+1)
t \right) ^{2}},
\end{eqnarray}
where $\rho_{_{0}}=3 H^2_{_0}$, in order to have
$a(t_{_0}=0)=a_{_0}>0$ and $H(t_{_0}=0)=H_{_0}>0$.

It is easy to verify that in this case the
metric~(\ref{INHOMOGENEOUS FRW}) becomes
\begin{eqnarray}\label{INHOMOGENEOUS FRW15}
ds^2=dt^2-a^2_{_{0}} \left(1+\frac{3}{2} H_{_0}(\omega+1) t
\right)^{4/(3(\omega+1))}  \times \nonumber \\ \left( \frac{dr^2}{1-
\left(\frac{r}{r_0}\right)^{-(1+\omega_r)/\omega_r}}+ r^2 (d\theta^2+sin^2 \theta d \varphi^2)\right), \nonumber \\
\end{eqnarray}
and the anisotropic and inhomogeneous energy density is given by
\begin{eqnarray}\label{rrrr}
\rho_{_{in}}(t,r)=-\frac{\left(\frac{r}{r_0}\right)^{-(1+3\omega_r)/\omega_r}}{r^2_0
\omega_r a^2_{_{0}} \left(1+\frac{3}{2} H_{_0}(\omega+1) t
\right)^{4/(3(\omega+1))} }.
\end{eqnarray}
The metric~(\ref{INHOMOGENEOUS FRW15})  represents an evolving
wormhole with a throat located at $r_0$ for $\omega_r<-1$ and
$\omega_r>0$~\cite{Cataldo15} and is asymptotically a flat FRW
universe.

It is well known that in general to keep a wormhole open exotic
matter with a negative energy density at the throat is
needed~\cite{Morris}. However, there are examples of evolving
wormholes satisfying the DEC in the whole
spacetime~\cite{Maeda,Cataldo15}, which implies that the energy
density is positive everywhere. In the studied here solutions the
branch with $\omega_r>0$ is characterized by a positive radial
pressure $p_r$, and negative $\rho_{_{in}}$ and $p_l$. But, there
are also wormholes which avoid the usual exotic matter requirements
for wormholes. The branch with $\omega_r < -1$ has a positive energy
density and lateral pressure, while the radial pressure is negative
and larger in magnitude than the energy density. Specifically, the
total matter content
\begin{eqnarray}\label{total energy}
\rho_{_{T}}(t,r)= \rho(t)+\rho_{_{in}}(t,r)
\end{eqnarray}
for these models is determined by Eqs.~(\ref{EDh}) and~(\ref{rrrr}).
Thus, for any value of the state parameters $\omega$ and
$\omega_r<0$ we have that always $\rho_{_{T}} \geq 0$. For
$\omega_r>0$ we can have in general time intervals where the total
energy is positive or negative. In effect, for $\omega>-1/3$ the
wormhole model starts with a positive total energy density (since
for a fixed value $r=const$ the isotropic component dominates over
the another one), then decreases till zero at certain $t_{0}$, and
becomes negative for $t> t_{0}$. For $\omega < -1/3$ the total
energy density starts negative, then increases till zero at certain
$t_{0}$, and becomes positive for $t> t_{0}$. If $\omega=-1/3$ both
components of total energy density~(\ref{total energy}) behave as
$\sim 1/a^2$, and the expansion occurs at a constant velocity.

In order to have accelerated (decelerated) expansion we must
consider $\omega<-1/3$ ($\omega>-1/3$). Notice that at $t=0$ we have
that the total energy density is given by
\begin{eqnarray}\label{condicion}
\rho_{_T}(0,r)=\rho_0-\frac{\left(\frac{r}{r_0}\right)^{-(1+3\omega_r)/\omega_r}}{r^2_0
\omega_r a^2_{_{0}} }.
\end{eqnarray}
We can see from Eq.~(\ref{condicion}) that if the values of
$\omega_r$ are constrained to be in the ranges $\omega_r<-1/3$ and
$\omega_r>0$, then the relation $\rho(0)>\rho_{_{in}}(0,r)$ is
fulfilled for $r>( -r_0^{1+1/\omega_r} \omega_r^{-1} a_0^{-2}
\rho_0^{-1})^{\omega_r/(1+3 \omega_r)}$. If the last inequality is
not fulfilled, then at $t=0$ the inhomogeneous component begins
dominating. If $-1/3 < \omega_r < 0$ we have that
$\rho_{_{in}}(0,r)> \rho(0)$ for $r>(-r_0^{1+1/\omega_r}
\omega_r^{-1} a_0^{-2} \rho_0^{-1})^{\omega_r/(1+3 \omega_r)}$.

\begin{figure}
\includegraphics[scale=0.3]{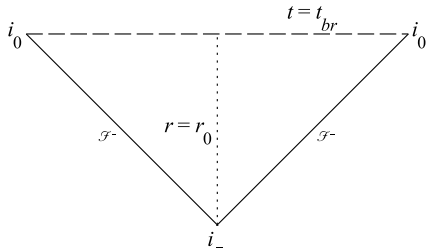}
\caption{Penrose diagram for an evolving wormhole with $\omega_r<-1$
and $\omega<-1$. The dotted line is the wormhole throat and the
dashed line represents the future Big Rip singularity.}
\label{penrose}
\end{figure}

\begin{figure*}
  \centering
  \subfigure[\, $r_0<r_c< r_*$]{\label{fig:gull}\includegraphics[width=0.3\linewidth]{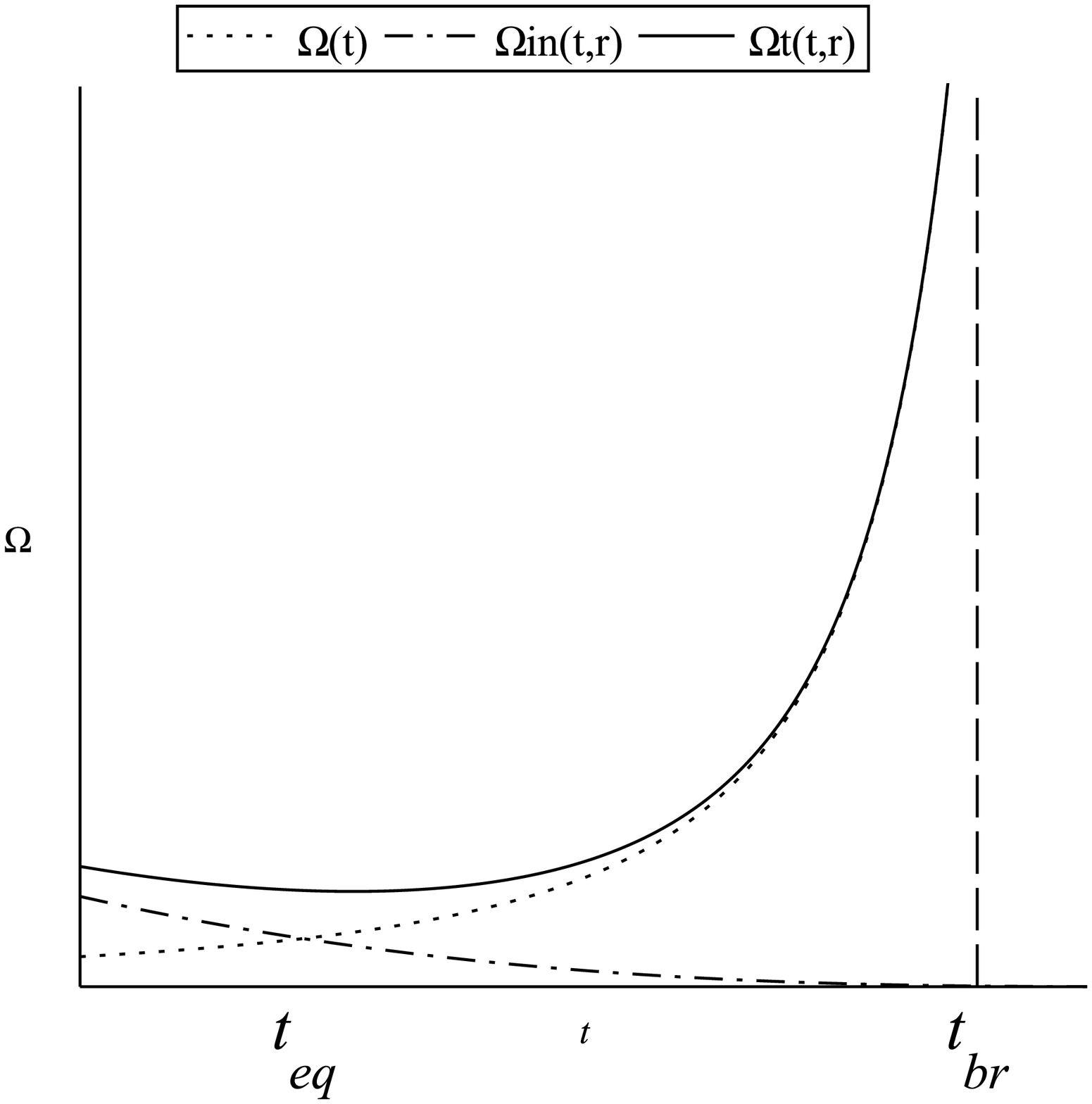}}
  \subfigure[\, $r_c>r_*$]{\label{fig:tiger}\includegraphics[width=0.3\linewidth]{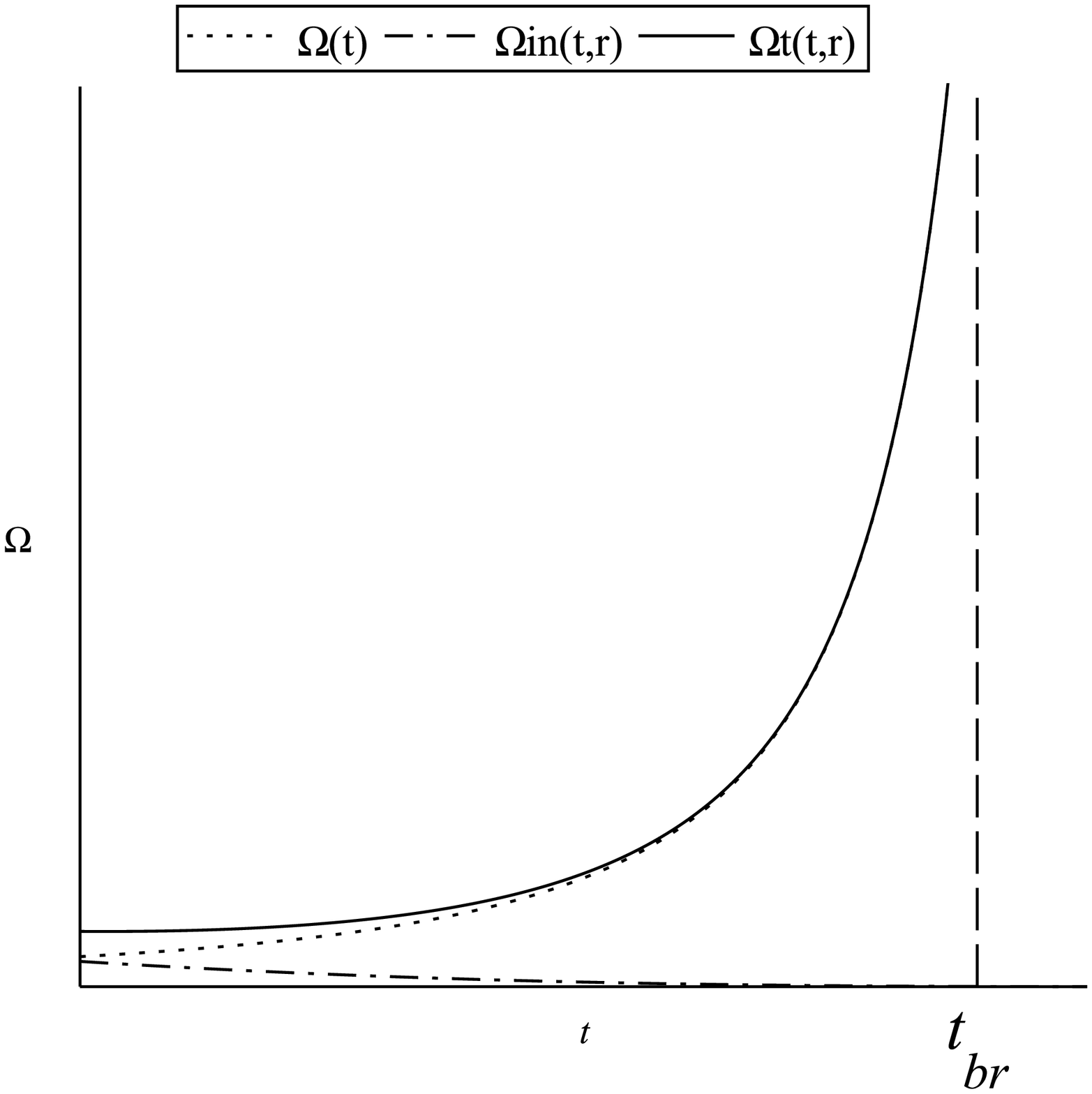}}
  \subfigure[\, $r_c\geq r_0 \, (r_* < r_0)$]{\label{fig:mouse}\includegraphics[width=0.3\linewidth]{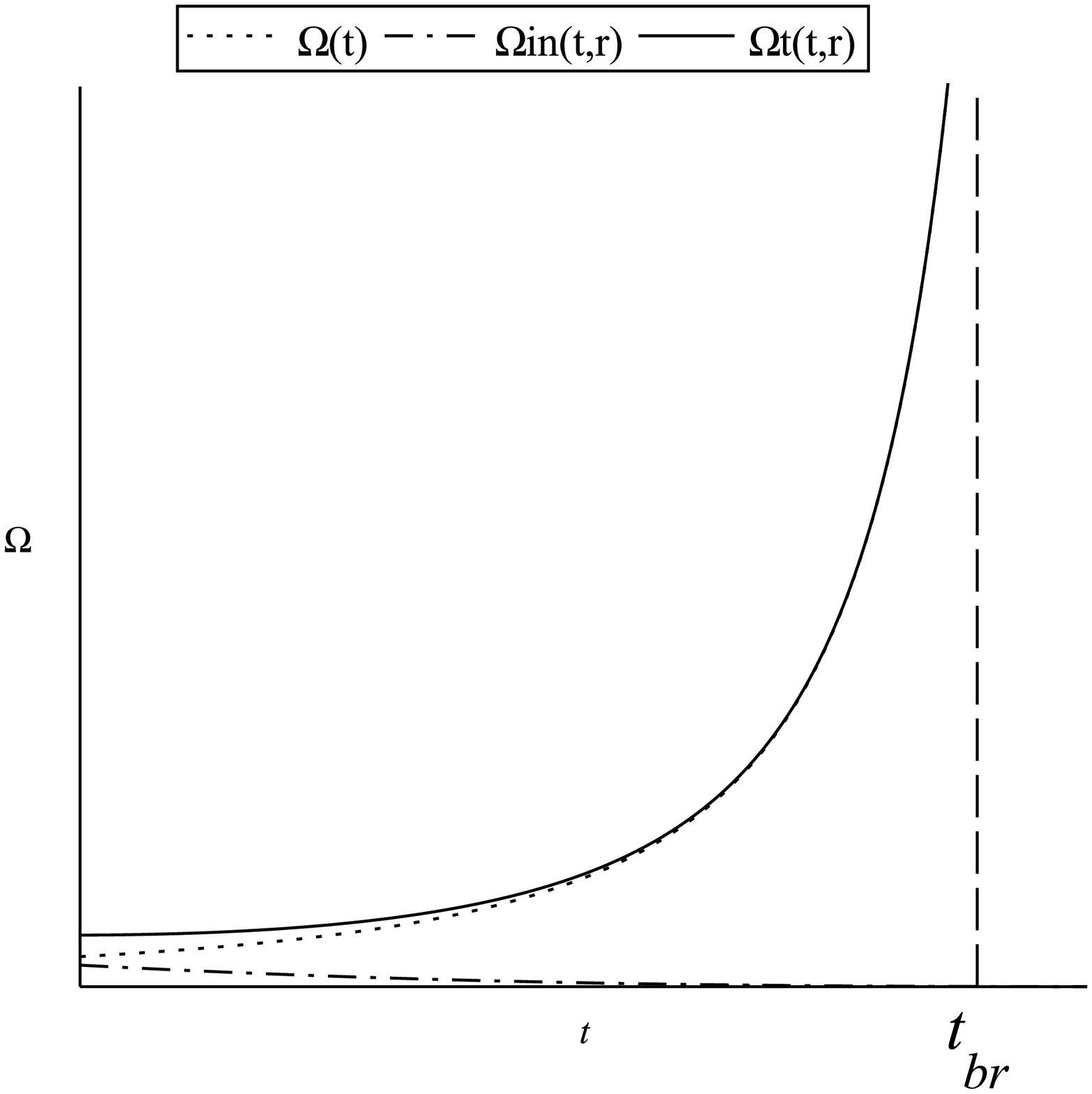}}
  \caption{These figures show, for wormhole models with $\omega_r<-1$ and $\omega<-1$, the qualitative behavior of the dimensionless
energy density for the isotropic component $\Omega(t)$ (dotted
line), anisotropic matter $\Omega_{in}(t,r_c)$ (dash-dotted line)
and the total matter content $\Omega_T(t,r_c)$ (solid line) as a
function of the time $t$, at a constant value of the radial
coordinate $r_c$. The comoving time varies from $0$ to the big rip
time $t_{_{br}}$. In Figs.~\ref{fig:gull} and~\ref{fig:tiger} are
plotted curves fulfilling the constraint~(\ref{otra condicion}): if
$r_0<r_c< r_*$ the anisotropic component dominates for $0 \leq
t<t_{_{eq}}$, while the isotropic component dominates over the
anisotropic one at the time interval $t_{_{eq}}<t<t_{_{br}}$
(Fig.~\ref{fig:gull}); on the other hand if $r_c>r_*$ the isotropic
component always dominates over the anisotropic for $0 \leq t <
t_{_{br}}$ (Fig.~\ref{fig:tiger}). In Fig.~\ref{fig:mouse} the
constraint~(\ref{otra condicion}) is not fulfilled, thus always the
isotropic component dominates over the anisotropic one for $0 \leq t
< t_{_{br}}$ and any constant value $r_c \geq r_0$. }
  \label{fig:animals}
\end{figure*}

\begin{figure*}
  \centering
  \subfigure[\, $r_0 < r_c < r^-_*$]{\label{fig:gull1}\includegraphics[width=0.3\linewidth]{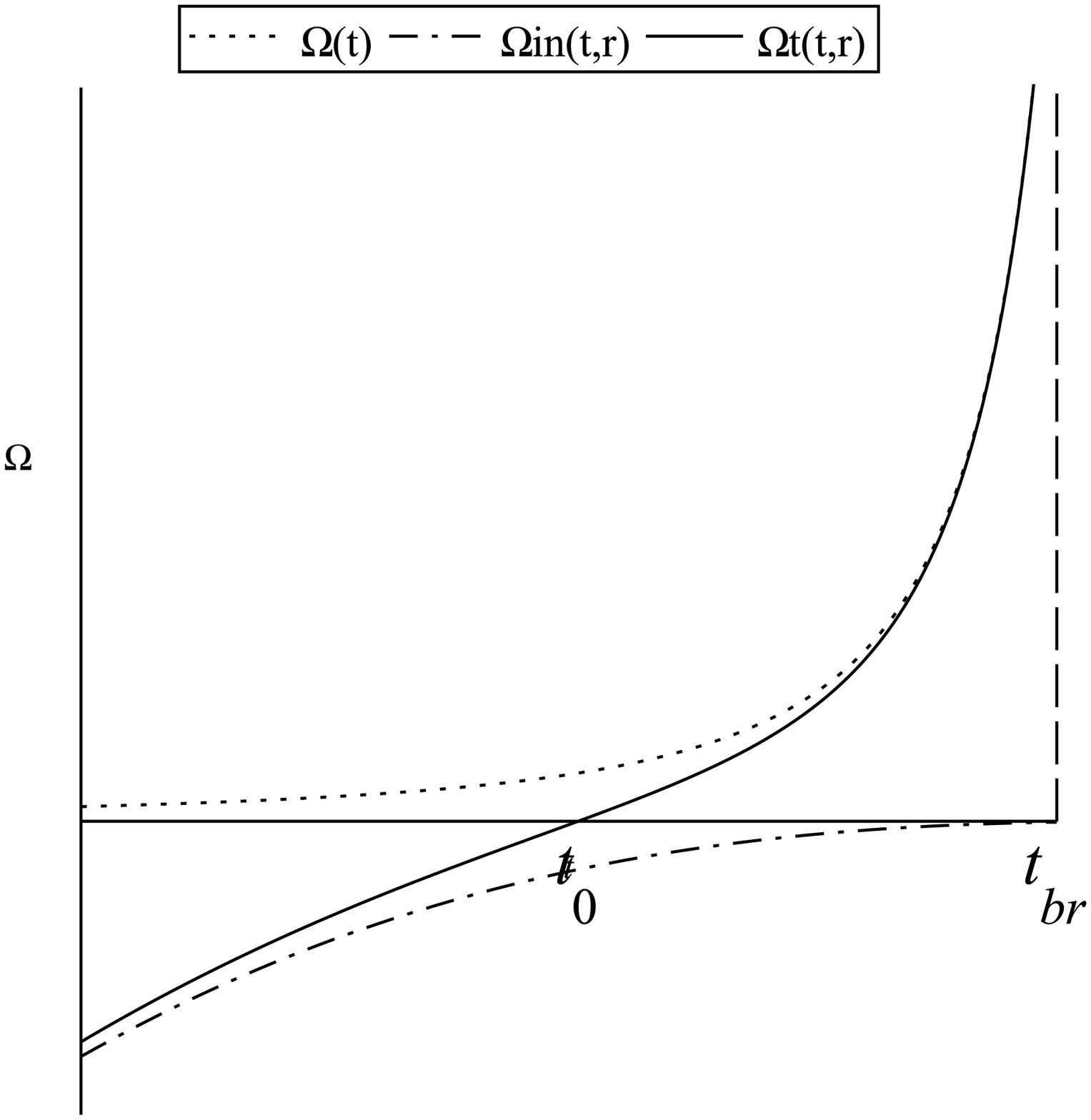}}
  \subfigure[\, $r_c>r^-_*$]{\label{fig:tiger1}\includegraphics[width=0.3\linewidth]{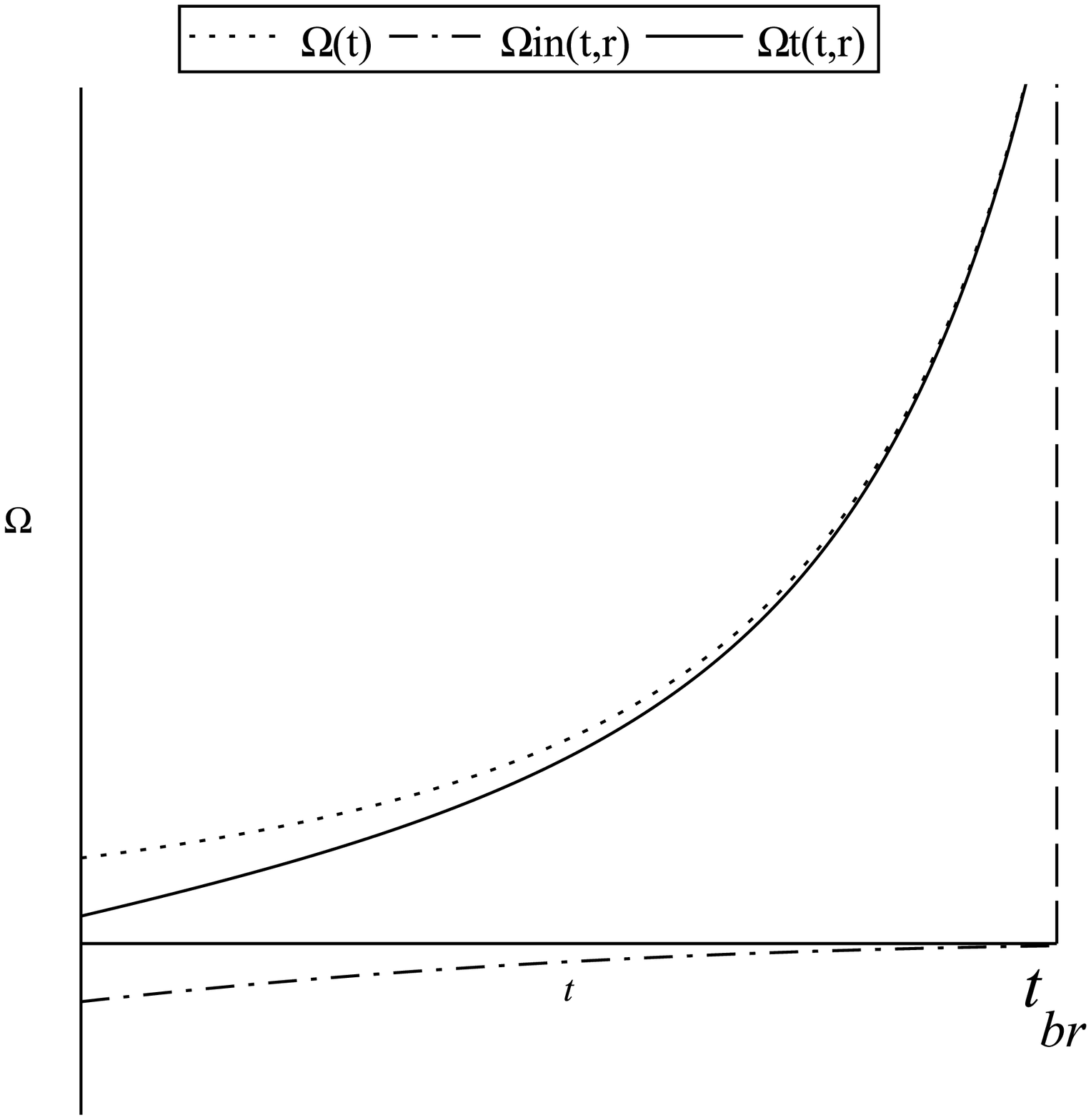}}
  \subfigure[$r_c\geq r^-_0 \, (r^-_*<r_0)$]{\label{fig:mouse1}\includegraphics[width=0.3\linewidth]{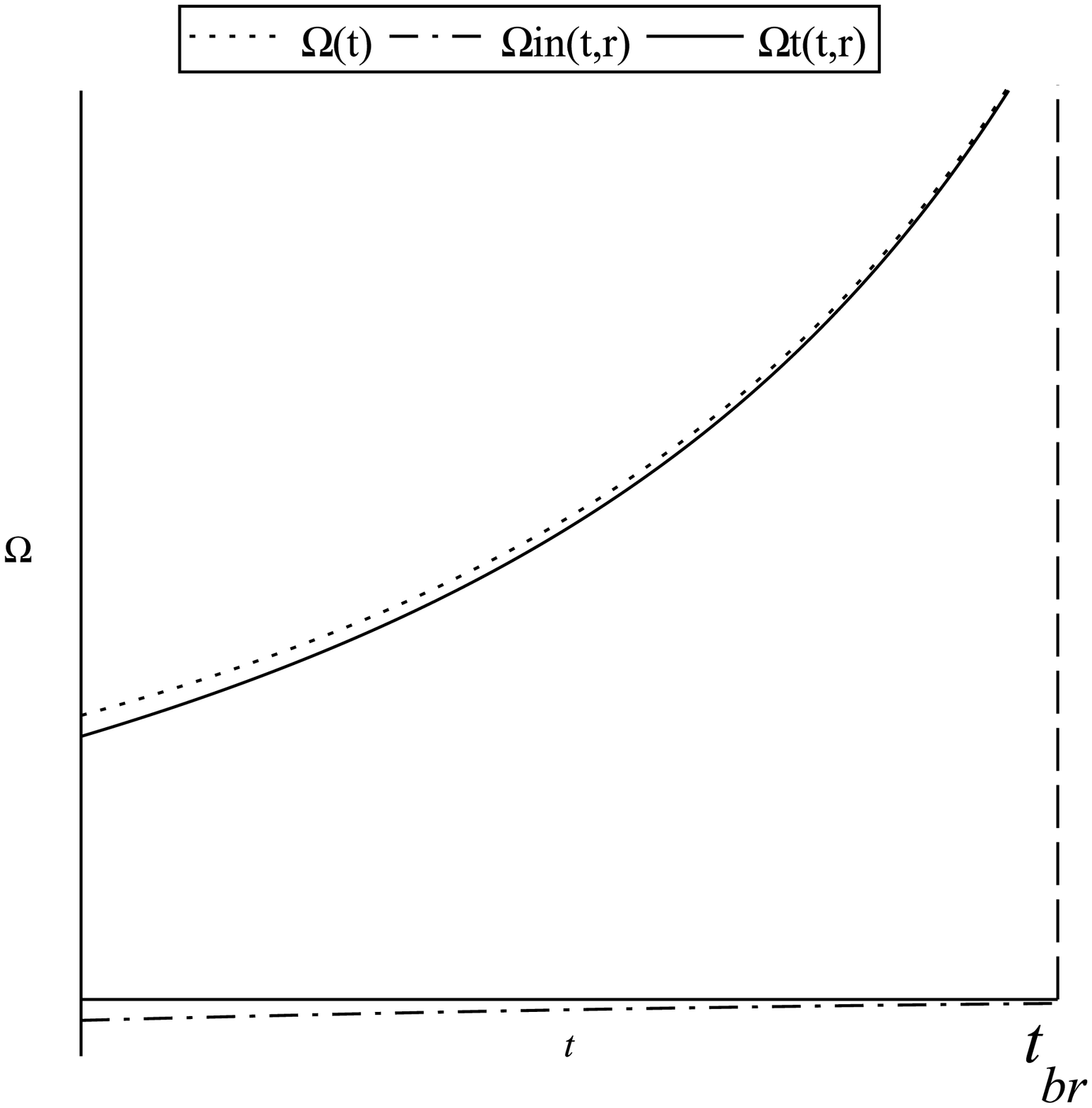}}
  \caption{These figures show, for wormhole models with $\omega_r>0$ and $\omega<-1$, the qualitative behavior of the dimensionless
energy density for the isotropic component $\Omega(t)$ (dotted
line), anisotropic matter $\Omega_{in}(t,r_c)$ (dash-dotted line)
and the total matter content $\Omega_T(t,r_c)$ (solid line) as a
function of the time $t$, at a constant value of the radial
coordinate $r_c$. The comoving time varies from $0$ to the big rip
time $t_{_{br}}$. In this case always the isotropic component
dominates over the anisotropic one. The total energy density
vanishes at $t=t_0$. For Fig.~\ref{fig:gull1} we have taken $0 < t_0
< t_{br}$, while for Figs~\ref{fig:tiger1} and~\ref{fig:mouse1} $t_0
< 0$. Here $r^-_*=r_0 (r_0^2 \omega_r a_0^2 \rho_0)^{-\omega_r/(1+3
\omega_r)}$. Note that in this case the anisotropic energy density
is always negative.}
  \label{fig:animals1}
\end{figure*}

For these  models it is useful to state that if the isotropic
perfect fluid satisfies the DEC, i.e. $-1< \omega <1$, we can
rescale the cosmological time so that $1+\frac{3}{2}
H_{_0}(\omega+1) t \longrightarrow  t$ and the solution for the
scale factor takes the standard form $a(t)=a_0
t^{{2/(3(\omega+1))}}$, with the isotropic energy density given by
$\rho(t)=\rho_0/t^2$.

In order to study these models during a phantom evolution we must
consider the values $\omega<-1$. In this case we have that
$1+\frac{3}{2} H_{_0} (\omega+1) t \equiv 1-\frac{3}{2} H_{_0}
|\omega+1| t$, so during the cosmic evolution this expression may
vanish at some value of time $t>0$. Thus we conclude that if $\omega
<-1$ we have a future singularity at a finite value of the proper
time
\begin{eqnarray}\label{singularity time}
t_{_{br}}=-\frac{2}{ 3 H_{_0} (\omega+1)} > t_{_{0}}=0,
\end{eqnarray}
since $ a(t) \longrightarrow \infty$, $ \rho(t) \longrightarrow
\infty$ and $p \longrightarrow - \infty$ at $t=t_{_{br}}$. By
considering only the behavior of scale factor and the isotropic
fluid we may conclude that this future singularity is of a Big Rip
type.

However we must also consider the presence of the inhomogeneous
matter component~(\ref{rrrr}). As we have stated above in this case
we have that $\rho_{_{in}}(t,r)>0$ for $\omega_r<0$, and
$\rho_{_{in}}(t,r)<0$ for $\omega_r>0$. For $\omega_r=-1/3$ this
matter component becomes homogeneous and isotropic one with
$p_{_r}=p_{_l}=-\rho_{_{in}}/3$. Note that for $\omega>-1$,
$\omega_r < -1/3$ and $\omega_r > 0$ we have that
$\rho_{_{in}}(t_c,r \longrightarrow \infty) \longrightarrow 0$ and
$\rho_{_{in}}(t \longrightarrow \infty,r_c) \longrightarrow 0$ for
constant values $t_c$ and $r_c$.

For $\omega < -1$, $\omega_r < -1/3$ and $\omega_r > 0$ we have that
at $t=0$ the anisotropic and inhomogeneous energy density starts to
evolve from the constant value $\rho_{_{in}}(0,r_c)$, where the
constant $r_c \geq r_0$, and then decreases, becoming zero at
$t=t_{_{br}}$.

As a consequence, for $t=t_{_{br}}$ and $\omega <-1$, we have that $
a(t) \longrightarrow \infty$, $ \rho(t) \longrightarrow \infty$, $p
\longrightarrow - \infty$, $\rho_{_{in}}(t,r) = 0$, $p_{_r}(t,r)
=p_{_l}(t,r) = 0$. This allows us to conclude that if we consider
the total matter content given by Eq.~(\ref{total energy}), for
$t=t_{_{br}}$ and $\omega <-1$, we have that $a(t) \longrightarrow
\infty$, $ \rho_{_T}(t,r) \longrightarrow \infty$, $p_{_{Tr}}(t,r)
\longrightarrow - \infty$, $p_{_{Tl}}(t,r) \longrightarrow -
\infty$, where $p_{_{Tr}}(t,r)=p(t)+p_{_{r}}(t,r)$ and
$p_{_{Tl}}(t,r)=p(t)+p_{_{l}}(t,r)$. In conclusion, the scale
factor, the total energy density and the total pressures blow up at
the finite time $t_{_{br}}$, so this finite-time future singularity
is of a Big Rip type. In Fig.~\ref{penrose} is shown a conformal
diagram of an evolving wormhole with $\omega_r<-1$ and a phantom
energy with $w < -1$.

Lastly, let us introduce the equilibrium time defined by the
condition $\rho(t_{_{eq}})=\rho_{_{in}}(t_{_{eq}},r)$, for wormholes
models with a positive total energy density and big rip. This allows
us to find $t_{_{eq}}$ as a function of the radial coordinate $r$:
\begin{eqnarray}\label{teq}
t_{_{eq}}=\frac{2}{3 H_{_0} (1+\omega)} \left[ \left(
-\frac{\left(\frac{r}{r_0} \right)^{-\frac{1+3
\omega_r}{\omega_r}}}{r_0^2 \omega_r a_0^2 \rho_0}
\right)^{\frac{-3(1+\omega)}{2(1+3\omega)}} -1 \right].
\end{eqnarray}

It becomes clear that in general $t_{eq}$ can take complex as well
as real values. In order to have always real values of the
equilibrium time the condition $\omega_r<0$ must be required. For
wormhole models with a positive total energy density and big rip the
equilibrium time is positive if $\omega_r<-1$, $\omega < - 1$ and
\begin{eqnarray}\label{otra condicion}
-\frac{\left(\frac{r}{r_0} \right)^{-\frac{1+3
\omega_r}{\omega_r}}}{r_0^2 \omega_r a_0^2 \rho_0} > 1 .
\end{eqnarray}
Thus, for $r_0<r< r_*$, where
\begin{eqnarray}\label{*}
r_*=r_0 (-r_0^2 \omega_r a_0^2 \rho_0)^{-\omega_r/(1+3 \omega_r)},
\end{eqnarray}
we have that $t_{_{eq}}>0$, and $t_{_{eq}}<0$ for $r>r_*$. Notice
that from Eqs.~(\ref{condicion}) and~(\ref{otra condicion}) we
obtain for the total energy density the equivalent constraint
$\rho_{_T}(0,r)> 2 \rho_0$, and from Eqs.~(\ref{singularity time})
and~(\ref{teq}) the condition $t_{eq}< t_{br}$ is automatically
fulfilled for $\omega_r<0$. By using Eqs.~(\ref{teq})-(\ref{*})
qualitative plots for wormholes defined by conditions $\omega_r<-1$
and $\omega_r>0$ are shown in Figs.~\ref{fig:gull}-\ref{fig:mouse}
and in Figs.~\ref{fig:gull1}-\ref{fig:mouse1} respectively.


Now we want to study evolving wormhole models with future
singularities and filled with a viscous phantom matter. This issue
will be addressed in the next section.

\section{Viscous evolving wormholes}

Let us now consider wormhole models with the isotropic and
homogeneous matter component described by a viscous phantom fluid.
The role of the dissipative processes has been extensively
considered in cosmology~\cite{Viscous Cosmology,CataldoLepe}, where
the study is done within the framework of the standard Eckart theory
of relativistic irreversible thermodynamics. Any dissipation process
in a FRW cosmology is scalar, and therefore may be modeled as a bulk
viscosity within a thermodynamical approach. The bulk viscosity
introduces dissipation by only redefining the effective pressure,
$P_{_{eff}}$, according to
\begin{eqnarray}\label{effecti}
P_{_{eff}}= p+\Pi= p-3 \xi H,
\end{eqnarray}
where $\Pi=\Pi(t)$ is the bulk viscous pressure, $\xi=\xi(t)$ is the
bulk viscosity coefficient and $H$ is the Hubble parameter.

In this case the Friedmann equations~(\ref{friedmann equation})
and~(\ref{FRWCEq}), with $k=0$ and $\kappa=8 \pi G=1$, take the form
\begin{eqnarray}\label{VFRW}
3 H^2=\rho+\Lambda, \\
\dot{\rho}+ 3 H (\rho+p+\Pi)=0. \label{ConsEq}
\end{eqnarray}
The violation of DEC is expressed by the relation $\rho+p+\Pi<0$.
This condition implies an increasing energy density of the isotropic
fluid filling the evolving wormhole, for a positive bulk viscosity
coefficient. The condition $\xi >0$ guaranties a positive entropy
production and, in consequence, no violation of the second law of
the thermodynamics~\cite{Pavon}.

We shall assume that the viscous component obeys the state
equation~(\ref{SEPF}), hence from Eq.~(\ref{effecti}) we have that
$P_{_{eff}}= \omega \rho -3 \xi H$. Thus from Eqs.~(\ref{VFRW})
and~(\ref{ConsEq}) we obtain the following evolution equation for H:
\begin{eqnarray}\label{Hpunto}
2\dot{H}+ 3 (\omega+1) H^{2}=3\xi H+(\omega+1)\Lambda.
\end{eqnarray}
From this equation we obtain for $\Lambda=0$ that
\begin{eqnarray}
H(t)=\frac{e^{\frac{3}{2}\int \xi(t) dt}}{C+\frac{3}{2} (\omega+1)
\int e^{\frac{3}{2}\int \xi(t) dt} \, dt}.
\end{eqnarray}
Thus for $\omega \neq -1$ the scale factor is given by
\begin{eqnarray}\label{scalefactor1}
a(t)= D \left(C+\frac{3}{2} \, (\omega+1) \int e^{\frac{3}{2} \,
\int \xi(t) dt} \, dt\right)^{2/(3 (\omega+1))}, \,\,\,\,\,
\end{eqnarray}
while for $\omega=-1$ it may be written as
\begin{eqnarray}\label{scalefactorpmenosrho}
a(t)= D e^{C\int e^{\frac{3}{2} \, \int \xi(t) dt} \, dt},
\end{eqnarray}
where $C$ and $D$ are integration constants. In general, for the
considered case, the solution may be written through $\xi(t)$ or
$a(t)$ because there are three independent equations for the four
unknown functions $a(t)$, $\rho(t)$, $\xi(t)$ and $p(t)$. In our
case we have written the solution through the bulk viscosity
$\xi(t)$. It is worth to mention that for a given $a(t)$ we can
write $H$ and then obtain the expressions for the energy density
from Eq.~(\ref{VFRW}) and the bulk viscosity from
Eq.~(\ref{Hpunto}).

It becomes clear that for $\xi=0$ and $\omega \neq -1$ we obtain
from Eq.~(\ref{scalefactor1}) the solution~(\ref{SFsinxi}) discussed
in the previous section, while for a vanishing bulk viscosity the
de~Sitter scale factor $a(t)=e^{H_{_{0}} t }$ is obtained for
$\rho=-p=const$.

On the other hand, note that from Eq.~(\ref{VFRW}) and
Eq.~(\ref{ConsEq}) we may write that
\begin{eqnarray}\label{1rr}
\frac{\ddot{a}}{a}=\dot{H}+H^2=-\frac{1}{6} \left(\rho + 3
P_{_{eff}} \right)+\frac{\Lambda}{3}.
\end{eqnarray}

Thus the condition for an expansion with constant velocity is given
by $\rho + 3 P_{_{eff}}=2 \Lambda$. By taking into account
Eq.~(\ref{effecti}) we may write
\begin{eqnarray}
\xi=\frac{1}{9 H} \left(\frac{}{} (1+3 \omega) \rho-2 \Lambda
\frac{}{} \right).
\end{eqnarray}
Note that for $\Lambda=0$, and by taking into account
Eq.~(\ref{VFRW}), we conclude that in order to have dynamic
wormholes expanding with constant velocity the bulk viscosity must
be given by
\begin{eqnarray}\label{constant expansion}
\xi=\frac{(1+3 \omega)}{3 \sqrt{3}} \rho^{^{1/2}}.
\end{eqnarray}
In this case we see that a necessary condition to have a positive
bulk viscosity coefficient is that $\omega> -1/3$.

Now we shall consider specific viscous phantom evolving wormhole
models.

\subsection{Wormhole models with constant bulk viscosity}
Let us now consider wormhole models with a vanishing cosmological
constant and a bulk viscosity given by
\begin{eqnarray}
\xi(t)=\xi_{_{0}}=const.
\end{eqnarray}
For $\omega \neq -1$, the Eq.~(\ref{scalefactor1}) allows us to
write the scale factor in the form
\begin{eqnarray}
a(t)=a_{_{0}} \left( 1+\frac{H_{_0}}{\xi_{_0}}\, (\omega+1) \left(
e^{\, 3 \, \xi_{_{0}}\, t/2}-1 \right)\right)^{2/(3 (\omega+1))},
\nonumber \\
\end{eqnarray}
from which we obtain that the homogeneous and isotropic energy
density is given by
\begin{eqnarray}\label{viscous energy1}
\rho(t)=  \frac{3\,{H}_{_0}^{2} e^{3 \xi_{_{0}} t}}{\left(1+
\frac{H_{_0}}{\xi_{_0}} \,(\omega+1) \left( e^{\, 3 \, \xi_{_{0}}\,
t/2}-1 \right)\right)^{2}}.
\end{eqnarray}
Then the anisotropic and inhomogeneous matter component takes the
following form:

\begin{eqnarray}\label{viscous inhomogeneous energy1}
\kappa \rho_{_{in}}(t,r)= \nonumber \\
-\frac{\left(\frac{r}{r_0}\right)^{-(1+3\omega_r)/\omega_r}}{r^2_0
\omega_r a^2_{_{0}} \left( 1+\frac{H_{_0}}{\xi_{_0}}\, (\omega+1)
\left( e^{\, 3 \, \xi_{_{0}}\, t/2}-1 \right)\right)^{4/(3
(\omega+1))}}.
\end{eqnarray}


As in the previous section, for $\omega < -1$ we have a future
singularity at a finite value of the comoving proper time
$t_{_{br}}$ since $ a(t) \longrightarrow \infty$, $ \rho(t)
\longrightarrow \infty$ and $p \longrightarrow - \infty$ at
$t_{_{br}}=\frac{2}{3 \xi_{_{0}}} \, \ln(1-\frac{\xi_{_0}}{H_{_0}
(\omega+1)}) > 0$. In this case at t=$t_{_{br}}$ the energy density
of the anisotropic matter threading the wormhole vanishes since if
$\omega_r>0$ or $\omega_r<-1$ we obtain that $
\rho_{_{in}}(t_{_{br}},r) = 0$ for any $r \geq r_0$.

If we consider the total energy density $\rho_{_T}$, given now by
Eqs.~(\ref{total energy}), (\ref{viscous energy1}) and~(\ref{viscous
inhomogeneous energy1}), we conclude that for $\omega<-1$ $\rho_{_T}
\longrightarrow \infty$, $p_{_{Tr}} \longrightarrow \infty$ and
$p_{_{Tl}} \longrightarrow \infty$ at $t=t_{_{br}}$. Thus, this
future singularity is characterized by diverging scale factor, total
energy density and total pressures, but with a well behaved bulk
viscosity, since $\xi$ is constant during all evolution.

Notice that for the $\omega=-1$ branch solution we obtain from
Eq.~(\ref{scalefactorpmenosrho}) that the scale factor is given by
\begin{eqnarray}
a(t)= D e^{\frac{2C e^{\frac{3}{2}\xi_0 t}}{3 \xi_0}}.
\end{eqnarray}
In this case the model is characterized by an accelerated expansion
and does not end in a future singularity.

\subsection{Accelerating wormhole models with $\xi \sim \rho^{1/2}$}

Another interesting example in this line is obtained for a bulk
viscosity given by $\xi =\alpha \rho^{1/2}$, where $\alpha$ is a
constant parameter. In this case, for any value of the state
parameter $\omega$ and $\Lambda=0$, the integration of
Eq.~(\ref{Hpunto}) allows us to write
\begin{eqnarray}\label{Hdet}
H=\frac{H_0}{1+\frac{3}{2}H_0 (\omega+1 -\sqrt{3}\alpha)t},
\end{eqnarray}
where $H_{0}=H(t=0)$. Thus, the scale factor becomes
\begin{eqnarray}\label{scalefactor}
a(t)=a_{0}\left(1+\frac{3}{2}H_0 (\omega+1 -\sqrt{3}\alpha)t
\right)^{\frac{2}{3(\omega+1 -\sqrt{3}\alpha)}},
\end{eqnarray}
with $a_{0}=a(t=0)$. The energy density of the isotropic component
takes the form
\begin{eqnarray}\label{rhoBV}
\rho(t)=\frac{3H^2_0}{(1+\frac{3}{2}H_0 (\omega+1
-\sqrt{3}\alpha)t)^2},
\end{eqnarray}
while the bulk viscosity and the energy density of the anisotropic
and inhomogeneous fluid are given by
\begin{eqnarray}\label{xiraizmedia}
\xi(t)=\frac{\alpha \sqrt{3} H_0}{1+\frac{3}{2}H_0 (\omega+1
-\sqrt{3}\alpha)t},
\end{eqnarray}
\begin{eqnarray}\label{rhoBVV}
\kappa \rho_{_{in}}(t,r)= \nonumber \\
-\frac{\left(\frac{r}{r_0}\right)^{-(1+3\omega_r)/\omega_r}}{r^2_0
\omega_r a^2_{_{0}} \left( 1+\frac{3}{2}H_0 (\omega+1
-\sqrt{3}\alpha)t \right)^{{\frac{4}{3(\omega+1
-\sqrt{3}\alpha)}}}},
\end{eqnarray}
respectively.

By demanding that
\begin{eqnarray}\label{constrain}
\sqrt{3}\alpha > \omega+1
\end{eqnarray}
the scale factor, the isotropic energy density and pressure blow up
to infinity at a finite time
\begin{eqnarray}\label{tbig}
t_{br}=\frac{2 H_{0}^{-1}}{3(\sqrt{3}\alpha -(\omega+1))}>0,
\end{eqnarray}
and then we have the occurrence of a future singularity. As in the
previous case, at t=$t_{_{br}}$ the energy density and pressures of
the anisotropic matter threading the wormhole vanishes for any $r
\geq r_0$.

By considering the total energy density $\rho_{_T}$, given now by
Eqs.~(\ref{total energy}), (\ref{rhoBV}) and~(\ref{rhoBVV}), we
conclude that for the constraint~(\ref{constrain}) we have that
$\rho_{_T} \longrightarrow \infty$, $p_{_{Tr}} \longrightarrow
\infty$ and $p_{_{Tl}} \longrightarrow \infty$ at $t=t_{_{br}}$.
Thus, this future singularity is characterized by diverging scale
factor, total energy density and total pressures, and diverging too
bulk viscosity~(\ref{xiraizmedia}), since it blows up at the time
$t_{br}$. In conclusion, in these viscous expanding wormhole models
the scale factor, the total energy density, the total pressures and
the bulk viscosity blow up at the finite time $t_{_{br}}$, so this
finite-time future singularity is of a Big Rip type.

\begin{figure*}
  \centering
  \subfigure[\, $\omega_r <-1$]{\label{fig:Vgull}\includegraphics[width=0.3\linewidth]{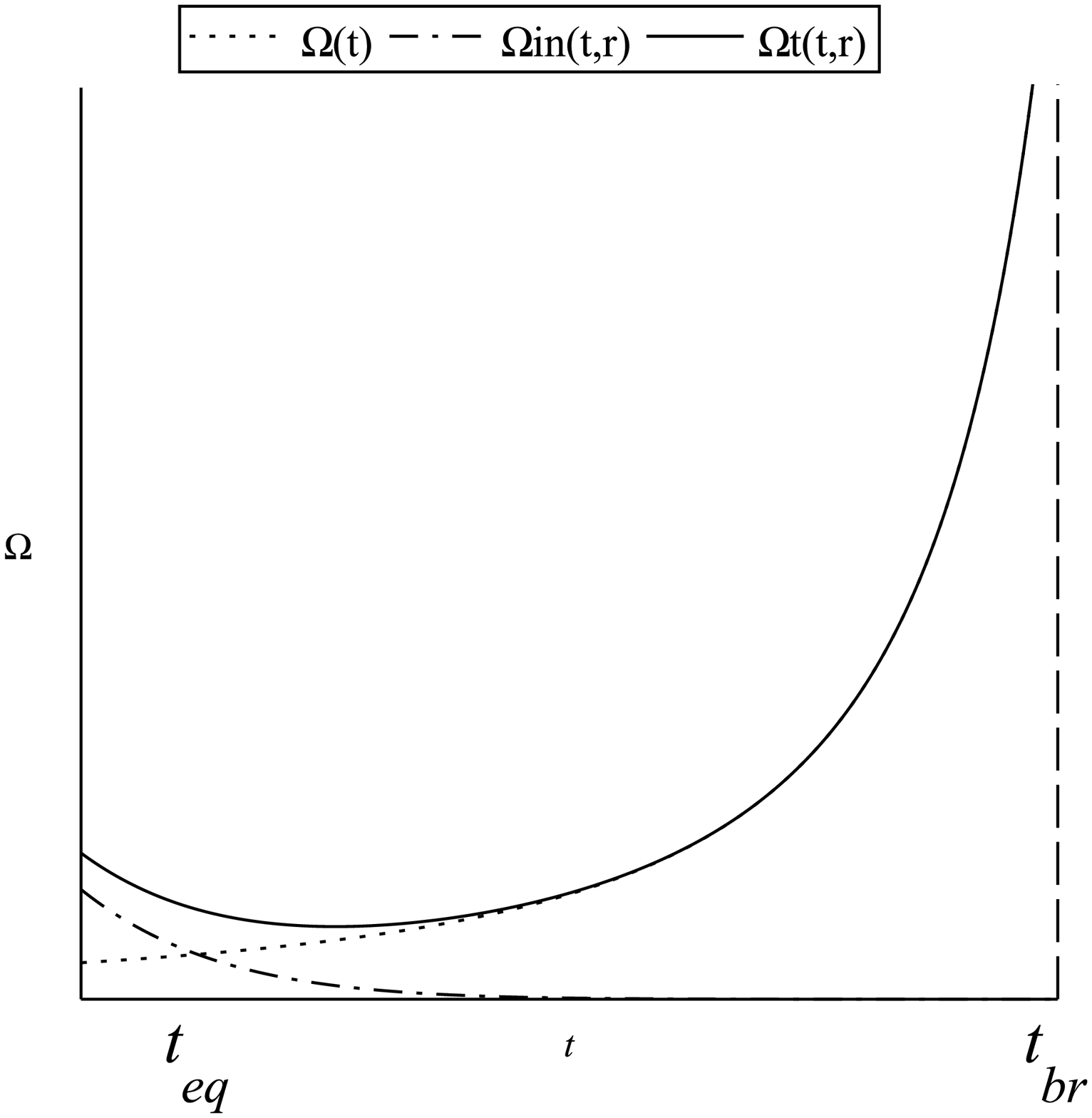}}
  \subfigure[\, $\omega_r>0$]{\label{fig:Vtiger}\includegraphics[width=0.3\linewidth]{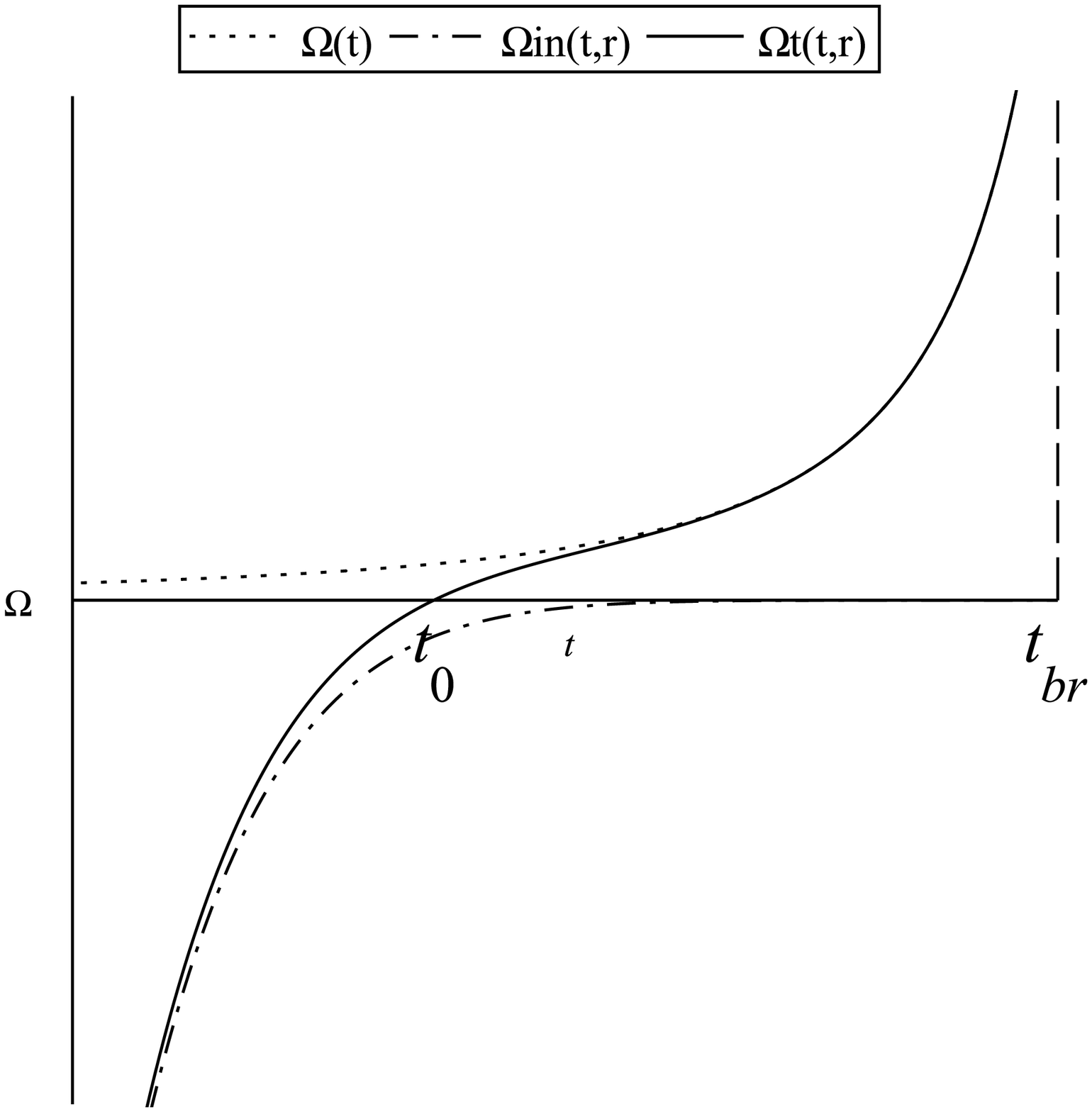}}
  \caption{These figures show, for viscous wormholes with $\xi \sim \rho^{1/2}$, the qualitative behavior of the dimensionless
energy density for the isotropic component $\Omega(t)$ (dotted
line), anisotropic matter $\Omega_{in}(t,r_c)$ (dash-dotted line)
and the total matter content $\Omega_T(t,r_c)$ (solid line) as a
function of the time $t$, at a constant value of the radial
coordinate $r_c$. The comoving time varies from $0$ to the big rip
time $t_{_{br}}$. For wormholes defined by $\omega_r <-1$, we can
have a similar behavior for all values of the state parameter
$\omega$, i.e. for $\omega<-1$, $-1<\omega<-1/3$ and $\omega>-1/3$,
as is shown in Fig.~\ref{fig:Vgull}. Here $t_{_{eq}}$ represents the
time at which $\Omega_{in}(t,r_c)=\Omega(t)$ and the big rip
singularity is present even if $\omega>-1$. For wormholes defined by
$\omega_r >0$ we have again a similar behavior for all values of the
state parameter $\omega$, i.e. for $\omega<-1$, $-1<\omega<-1/3$ and
$\omega>-1/3$, as is shown in Fig.~\ref{fig:Vtiger}. Note that in
this case the anisotropic energy density is always negative. The big
rip singularity is also present even if $\omega>-1$ and $t_0$
represents the time at which $\Omega_T(t,r_c)=0$.}
  \label{fig:animals15}
\end{figure*}

It is interesting to note that the constraint~(\ref{constrain})
implies that if the bulk viscosity is positive, i.e. $\alpha>0$, we
can have a future singularity also for $\omega \geq -1$. Thus we
have a big rip singularity not only for viscous phantom energy, but
also for viscous dark energy, and even for standard viscous matter
(see Figs.~\ref{fig:Vgull} and~\ref{fig:Vtiger}). Clearly all these
models have an accelerated expansion. It is easy to show that in
order to have models expanding with constant velocity we must
require that $\frac{2}{3(\omega+1 -\sqrt{3}\alpha)}=1$. This implies
that $\alpha=(1+3 \omega)/3\sqrt{3}$ in agreement with
Eq.~(\ref{constant expansion}).
 \label{Figura1 15}


\section{Conclusions and further comments}
This paper deals with inhomogeneous and anisotropic spacetimes,
ending in a future singularity at a finite value of the proper time,
and filled with an inhomogeneous and anisotropic fluid and another
isotropic and homogeneously distributed super-quintessence fluid.
The studied solutions describe evolving wormholes for which the rate
of expansion is determined by the phantom energy, while the
inhomogeneous and anisotropic component threads and sustains the
wormhole. The main purpose of this work is to present analytic
wormhole models exhibiting a Big Rip during its evolution. Three
independent cosmological models are explored.

In the first model it turns out that, the isotropic and homogeneous
component is a barotropic phantom fluid. For these evolving
wormholes the scale factor, the total energy density and the total
pressures blow up at a finite proper time, so this finite-time
future singularity is of a Big Rip type. In the second wormhole
configuration the isotropic and homogeneous component is a viscous
phantom fluid with a constant bulk viscosity. Now, the future
singularity is characterized by diverging scale factor, total energy
density and total pressures, but with a well behaved bulk viscosity,
due to the constant character of $\xi$ during all evolution. In the
latter model the isotropic and homogeneous component is a viscous
phantom fluid with a bulk viscosity of the form $\xi \sim
\rho^{1/2}$. For these viscous dynamic wormholes the scale factor,
the total energy density, the total pressures and the bulk viscosity
blow up at a finite proper time, so this future singularity is also
of a Big Rip type. It must be added that this third model allows us
to consider Big Rip wormholes not only for a viscous
super-quintessence energy. Effectively, if $\xi>0$, we can have a
Big Rip also for viscous dark energy (i.e. for $-1< \omega < -1/3$),
and even for standard viscous matter (i.e. for $\omega>-1/3$). If
$\alpha<0$ the Big Rip is avoided. However in this case the bulk
viscosity becomes negative and, in consequence, the second law of
the thermodynamics is not fulfilled.

Notice that, in all considered here solutions the mixed component of
the energy-momentum tensor $T_{tr}$ vanishes. This means that there
is no radial energy flow and no accretion onto the wormhole of
phantom energy from the cosmic fluid. Thus the mechanism by which
the big trip could be achieved is out of the possibilities for these
wormhole models~\cite{Faraoni}.

Lastly, we want to state that all obtained here results on future
singularities are applicable also for flat FRW cosmological models,
since all the discussed wormhole solutions are asymptotically flat
FRW cosmologies.

\section{Acknowledgements}
This work was supported by CONICYT through Grant FONDECYT N$^0$
1080530 and by the Direcci\'on de Investigaci\'on de la Universidad
del Bio-B\'\i o through grants N$^0$ DIUBB 121007 2/R and N$^0$
GI121407/VBC (MC).

\end{document}